\documentclass[a4paper,11pt]{article}
\usepackage{jinstpub} 
\usepackage{todonotes}
\usepackage{multirow}
\usepackage{subfigure}
\usepackage{siunitx}
\usepackage{placeins}
\usepackage[T5,T1]{fontenc}

\DeclareSIUnit{\pe}{PE}

\title{\boldmath Acceptance Tests of more than 10\,000 Photomultiplier Tubes for the multi-PMT Digital Optical Modules of the IceCube Upgrade}

\author[16]{R. Abbasi,}
\author[64]{M. Ackermann,}
\author[17]{J. Adams,}
\author[39,a]{S. K. Agarwalla,}
\author[11]{J. A. Aguilar,}
\author[21]{M. Ahlers,}
\author[22]{J.M. Alameddine,}
\author[43]{N. M. Amin,}
\author[41]{K. Andeen,}
\author[25]{G. Anton,}
% ...
\affiliation[11]{Universit{\'e} Libre de Bruxelles, Science Faculty CP230, B-1050 Brussels, Belgium}
% ...
\affiliation[16]{Department of Physics, Loyola University Chicago, Chicago, IL 60660, USA}
\affiliation[17]{Dept. of Physics and Astronomy, University of Canterbury, Private Bag 4800, Christchurch, New Zealand}
% ...
\affiliation[21]{Niels Bohr Institute, University of Copenhagen, DK-2100 Copenhagen, Denmark}
\affiliation[22]{Dept. of Physics, TU Dortmund University, D-44221 Dortmund, Germany}
% ...
\affiliation[25]{Erlangen Centre for Astroparticle Physics, Friedrich-Alexander-Universit{\"a}t Erlangen-N{\"u}rnberg, D-91058 Erlangen, Germany}
% ...
\affiliation[39]{Dept. of Physics and Wisconsin IceCube Particle Astrophysics Center, University of Wisconsin{\textemdash}Madison, Madison, WI 53706, USA}
% ...
\affiliation[41]{Department of Physics, Marquette University, Milwaukee, WI 53201, USA}
% ...
\affiliation[43]{Bartol Research Institute and Dept. of Physics and Astronomy, University of Delaware, Newark, DE 19716, USA}
% ...
\affiliation[64]{Deutsches Elektronen-Synchrotron DESY, Platanenallee 6, D-15738 Zeuthen, Germany}
\affiliation[a]{also at Institute of Physics, Sachivalaya Marg, Sainik School Post, Bhubaneswar 751005, India}
% ...
\collaboration{\includegraphics[height=17mm]{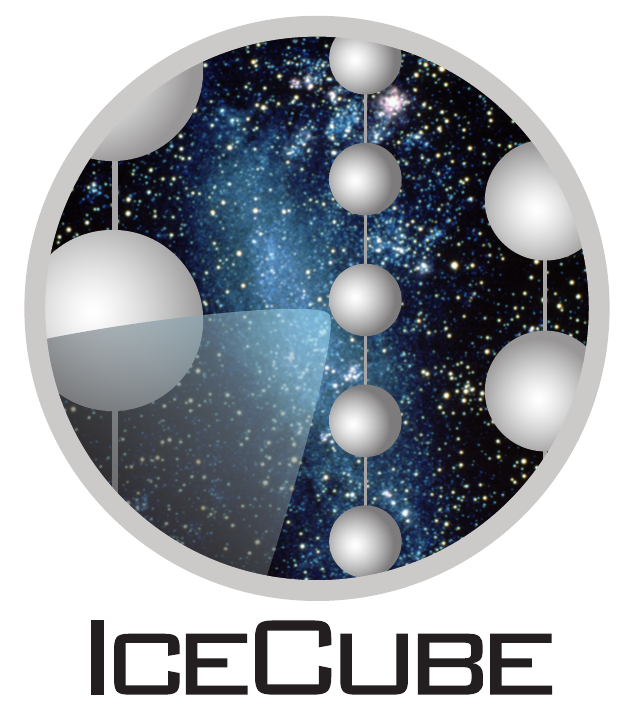}\\[6pt]The IceCube Collaboration}
\emailAdd{analysis@icecube.wisc.edu}

\abstract{More than \num{10000} photomultiplier tubes (PMTs) with a diameter of \SI{80}{\milli\meter} will be installed in multi-PMT Digital Optical Modules (mDOMs) of the IceCube Upgrade. These have been tested and pre-calibrated at two sites. A throughput of more than \num{1000} PMTs per week with both sites was achieved with a modular design of the testing facilities and highly automated testing procedures. The testing facilities can easily be adapted to other PMTs, such that they can, e.g., be re-used for testing the PMTs for IceCube-Gen2. Single photoelectron response, high voltage dependence, time resolution, prepulse, late pulse, afterpulse probabilities, and dark rates were measured for each PMT. We describe the design of the testing facilities, the testing procedures, and the results of the acceptance tests.}


\arxivnumber{2404.19589} 

\begin{document}
    
    \maketitle
    \flushbottom
    
    \section{Introduction}\label{sec:intro}

\begin{figure}[htbp]
    \centering
    \subfigure[mDOM technical drawing]{
        \includegraphics[width=0.49\textwidth]{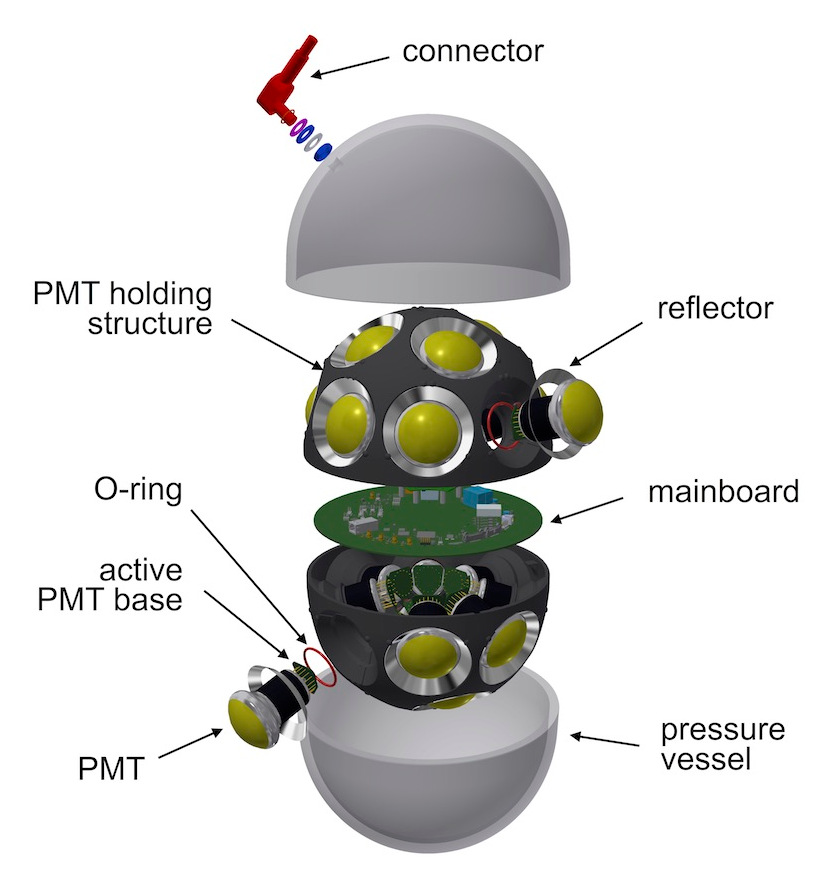}
    }
    \subfigure[mDOM PMT]{
        \includegraphics[width=0.45\textwidth]{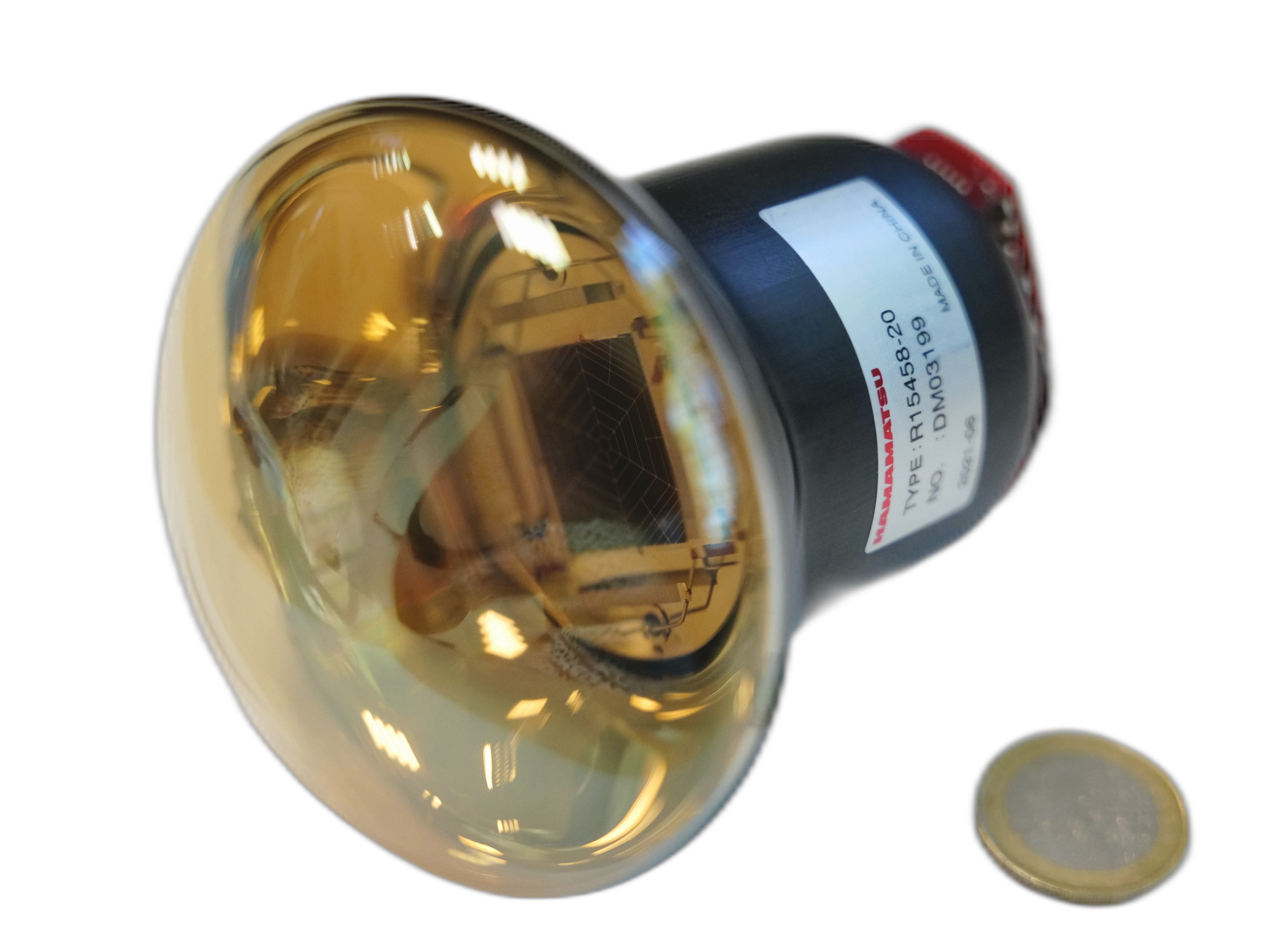}
    }
    \caption{Left: Technical drawing of an mDOM. The mDOM houses \num{24} PMTs in two hemispheres with an outer diameter of \SI{356}{\milli\meter}. The PMTs are mounted to a holding structure and are surrounded by reflector rings. Right: Hamamatsu Photonics K.K. R15458-20 Photomultiplier Tube for the mDOM with a coin for scale. The PMT has a bialkali photocathode with a diameter of \SI{80}{\milli\meter}; the window is made of borosilicate glass.}
    \label{fig:pmt}
\end{figure}

The IceCube Neutrino Observatory has been operating at the South Pole for more than a decade.
IceCube has discovered an all-sky extragalactic flux~\cite{diffuse_2013}, identified two active galaxies as likely sources~\cite{IceCube_NGC1068, IceCube_TXS}, and observed neutrinos from the galactic plane~\cite{IceCube_Galactic_Plane}.
It comprises \num{5160} optical sensors embedded into the glacial ice sheet at the South Pole, arranged on \num{86} vertical detector strings~\cite{IceCube_detector_paper}.
The IceCube Upgrade~\cite{IceCube_Upgrade_ICRC_2019} will dramatically improve measurements of neutrino oscillations with atmospheric neutrinos, provide improved calibration of the optical properties of the glacial ice, and act as a testbed for the next generation of optical modules designed for IceCube-Gen2~\cite{IceCube_Gen2_white_paper}.
Seven new detector strings of instrumentation will be deployed into the South Pole ice sheet in the 2025/2026 Antarctic Field season.
The multi-PMT Digital Optical Module (mDOM) is one of two main optical sensor types deployed in the IceCube Upgrade; the other is the D-Egg~\cite{D-Egg_paper}.
The deployment includes \num{405} mDOMs with \num{24} Photomultiplier Tubes (PMTs) each.
Including spares, this results in the need to test more than \num{10000} PMTs for compliance with performance requirements.
Figure~\ref{fig:pmt} shows a technical drawing of an mDOM and one of the PMTs, of type R15458-02, manufactured by Hamamatsu Photonics K.K.~\cite{R15458-02}.

The PMTs are operated with negative high voltage, generated by an active base on the back of each PMT with a \num{13}-stage Cockroft-Walton generator~\cite{cockroft_walton}.
This approach was chosen because of power requirements and limited space inside the mDOM~\cite{mDOM_ICRC_2021}.
The PMTs have ten dynodes; the first Cockroft-Walton stage sets the voltage $U_\mathrm{dy10}$ between the anode and dynode 10, the following nine stages connect to successive dynodes, and the combined last three stages set the voltage between the cathode and dynode 1~\cite{mDOM_ICRC_2021}.
The voltage of each stage falls off progressively towards the photocathode, resulting in a full photocathode-to-anode voltage of $U_\mathrm{full} = 12 \cdot U_\mathrm{dy10}$, where $U_\mathrm{dy10}$ is internally regulated to a value configured via a UART interface~\cite{mDOM_ICRC_2021}.
The active bases are designed to reach full voltages $U_\mathrm{full}\leq\SI{1.5}{\kilo\volt}$ and were soldered to the back of the PMTs by the manufacturer.
The PMT signals are provided to digitizers on the mDOM mainboards via a \SI{50}{\ohm} coax cable~\cite{mDOM_ICRC_2021}.

This paper describes the design and commissioning of two testing facilities, illustrates the measurement procedures and the automation strategy, and concludes with the results of the performed testing campaign.

    \section{Objective and Performance Requirements}\label{sec:task}

The performance requirements for the mDOM PMTs are listed in table~\ref{tab:specs}; they are largely based on requirements for PMTs in KM3NeT~\cite{Aiello_2018, Elorrieta_2019}. The requirements include specifications on the temperatures the PMTs can be operated and stored at without impact on their performance, on the supply voltage to reach an amplification gain of $g_\mathrm{target}=\SI{5e6}{\relax}$, on their sensitivity, pulse characteristics, timing characteristics, and on background rates from dark noise and signal-correlated pulses.
Some requirements were only tested on a few tens of PMTs separately from the testing campaign described here because of the technical limitations of the later-described testing facilities.
E.\,g., the size of suitable light sources at a wavelength of $\lambda=\SI{325}{\nano\meter}$ was not compatible with the design of our light source system.

\begin{table}[htbp]
    \centering
    \caption{Requirements for mDOM PMT Performance Parameters; they are largely based on requirements for PMTs in KM3NeT~\cite{Aiello_2018, Elorrieta_2019}. The requirements marked as tested are performed on all PMTs. The other requirements have been tested separately from the primary testing campaign for a few tens of PMTs.}
    \begin{tabular}{@{}c|c|c}
        Metric & Specification & Tested \\\hline\hline
        Operational Temperature & \SI{-45}{\degreeCelsius} to \SI{30}{\degreeCelsius} & \\
        Storage Temperature & \SI{-60}{\degreeCelsius} to \SI{50}{\degreeCelsius} & \\\hline
        Cathode Voltage @ Gain \SI{5e6}{\relax} & \SI{950}{\volt} to \SI{1350}{\volt} & \checkmark \\\hline
        & >~\SI{7}{\percent} @ \SI{325}{\nano\meter} &  \\
        Quantum Efficiency & >~\SI{25}{\percent} @ \SI{380}{\nano\meter} & \checkmark \\
        & >~\SI{15}{\percent} @ \SI{500}{\nano\meter} & \checkmark \\\hline
        Mean SPE Amplitude & >~\SI{6}{\milli\volt} @ Gain \SI{5e6}{\relax} and \SI{50}{\ohm} output impedance & \\
        Transit-Time-Spread ($\sigma$) & <~\SI{2.0}{\nano\second} with full frontal illumination @ SPE level & \checkmark \\
        Rise Time (10\% to 90\%) & >~\SI{1.0}{\nano\second} and <~\SI{5.0}{\nano\second} with \SI{100}{\pe} pulse &  \\
        SPE Charge Resolution ($\sigma$) & <~\SI{0.7}{\pe} & \checkmark \\
        Peak-to-Valley-Ratio & >~\SI{2.0}{\relax} & \checkmark \\
        Amplitude Linearity (within \SI{10}{\percent}) & >~\SI{100}{\pe} &  \\\hline
        Prepulses [\SI{-20}{\nano\second},\,\SI{-10}{\nano\second}] & <~\SI{1}{\percent} @ \SI{0.2}{\pe} threshold and \SI{-20}{\degreeCelsius} & \checkmark \\
        Late Pulses [\SI{15}{\nano\second},\,\SI{80}{\nano\second}] & <~\SI{5}{\percent} @ \SI{0.2}{\pe} threshold and \SI{-20}{\degreeCelsius} & \checkmark \\
        Afterpulses [\SI{100}{\nano\second},\,\SI{12}{\micro\second}] & <~\SI{15}{\percent} @ \SI{0.2}{\pe} threshold and \SI{-20}{\degreeCelsius} & \checkmark \\\hline
        \multirow{2}{*}{Dark Rate} & <~\SI{150}{\hertz} @ \SI{0.2}{\pe} threshold, $t_\mathrm{dead}=\SI{100}{\nano\second}$, & \multirow{2}{*}{\checkmark}\\
        & @\SI{-20}{\degreeCelsius} after \SI{5}{\hour} in dark
    \end{tabular}
    \label{tab:specs}
\end{table}

Since removing a faulty PMT from a fully integrated mDOM is practically impossible, the PMTs had to be tested prior to their integration.
This implied a high-throughput testing procedure to avoid slowing down the mDOM production.
This high throughput was achieved by high parallelization and automation within each of the two testing sites.

To mitigate potentially increased mDOM dark noise rates due to electric potential differences between neighboring PMT photocathodes~\cite{munland_phd}, we aim to minimize the photocathode voltage variance within each mDOM.
Therefore, in addition to acceptance testing, a preliminary PMT amplification gain calibration is performed, and the PMTs are sorted into six classes according to their nominal supply voltage for a gain of $g_\mathrm{nom} = \SI{5e6}{\relax}$.
All test results are stored in a central database of the IceCube Upgrade project for later use.

    \section{The Testing Facilities}\label{sec:facility}

\begin{figure}[htbp]
    \centering
    \includegraphics[width=1.0\textwidth]{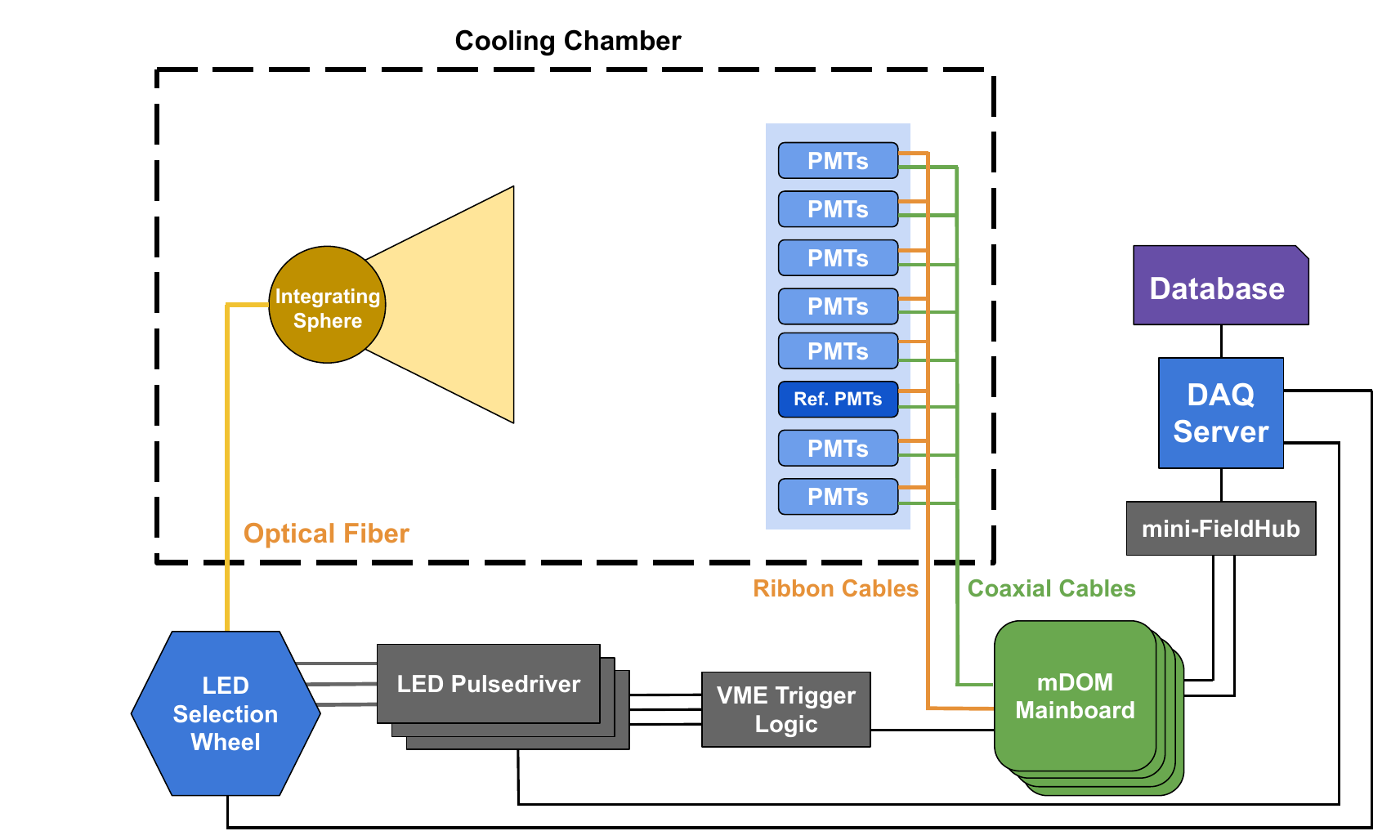}
    \caption{Block diagram of the testing facility. The PMTs are mounted in a rack inside a cooling chamber. They are illuminated by a polytetrafluoroethylene (PTFE) integrating sphere~\cite{POCAM_paper}, connected to a multi-wavelength light source system. Four mDOM mainboards~\cite{mDOM_mainboard_ICRC2023} connected to a control server comprise the Data Acquisition (DAQ) system.}
    \label{fig:facility_block_diagram}
\end{figure}

The design of the testing facilities has been optimized to perform all tests within fast and highly automated testing cycles.
Figure~\ref{fig:facility_block_diagram} shows a block diagram of the testing facility.
The central infrastructure is a light-tight, temperature-controlled dark room within which \num{92} PMTs are tested simultaneously at \SI{-20}{\celsius}~\cite{icrc2021_pmt_testing}.
The PMTs are mounted in a central rack and are illuminated by a multi-wavelength light source system.
Four reference PMTs are used to monitor the illumination levels and long-term stability of the test results~\cite{joelle-thesis}.
The Data Acquisition (DAQ) system comprises four mDOM mainboards~\cite{mDOM_mainboard_ICRC2023} connected to a control server that stores all data.
All active electronic components and the light source system are outside the cooling chamber.
This reduces performance changes due to temperature and humidity variations.
Special care was taken to make the cooling chamber airtight to avoid condensation from warm air rushing in.
During the final production of the mDOMs, the modules are filled with dry nitrogen.
Temperatures and humidities inside and outside the chamber were monitored continuously.

\subsection{PMT Mounting}

\begin{figure}[htbp]
    \centering
    \includegraphics[width=.95\textwidth]{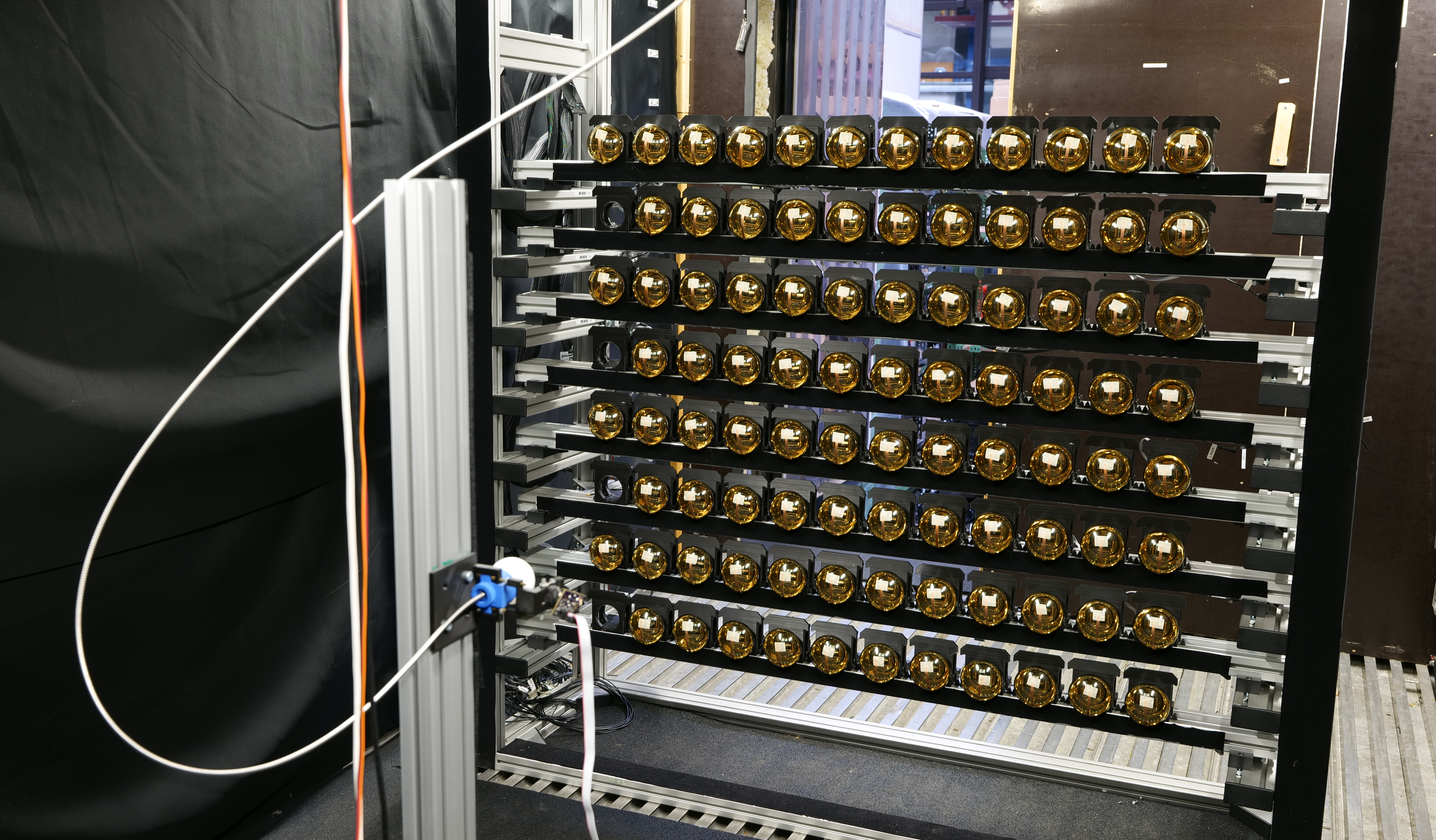}
    \caption{Frontside of the PMT Rack in the testing facility. \num{92} PMTs are mounted; the unoccupied channels are used for synchronization with the light source system. The optical fiber and PTFE integrating sphere are visible and centered in front of the PMT rack. The secondary door in the background is used as a humidity buffer.}
    \label{fig:facility_pmt_rack}
\end{figure}

The PMTs are mounted parallelly in a rack inside the cooling room.
Figure~\ref{fig:facility_pmt_rack} shows the PMT rack from the front with all PMTs for a test cycle mounted.
The PMT rack has eight rows, each equipped with an aluminum mounting rail with twelve PMTs mounted using 3D-printed holders.
The mounting of the rails in the rack and of the PMT holders on the rails is designed such that the structures can easily be adapted to different geometries, i.e., to test other PMT models~\cite{halve_phd}.
The mounting system is modular such that PMTs for the next testing cycle can already be mounted onto a second set of rails during ongoing tests, and only the rails need to be exchanged and reconnected inside the cooling room.

\begin{figure}[htbp]
    \centering
    \includegraphics[width=.9\textwidth]{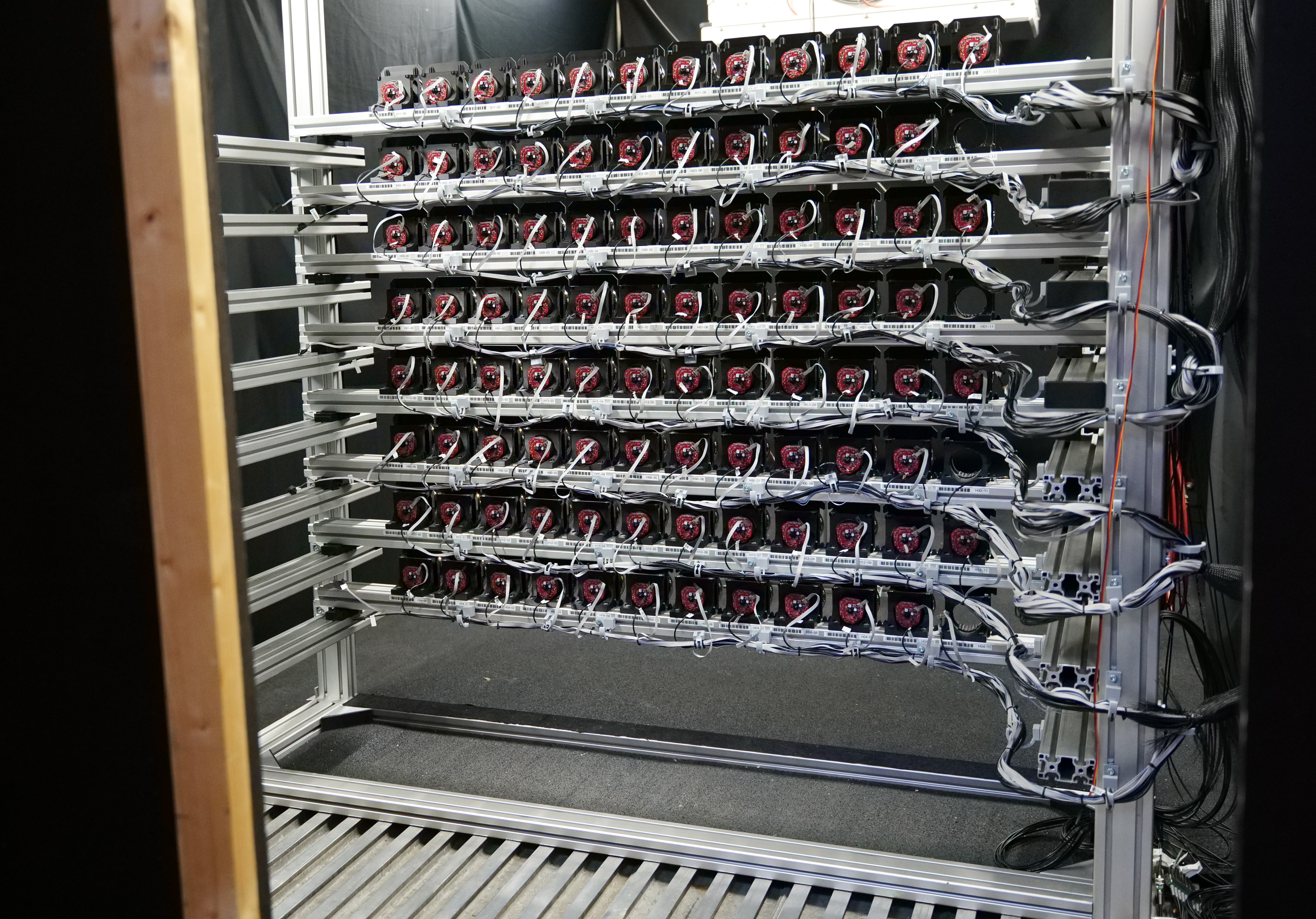}
    \caption{Backside of the PMT rack inside the testing facility. Each PMT is connected to an mDOM mainboard with a coaxial and ribbon cable. Each PMT position is marked with a barcode.}
    \label{fig:facility_cabling}
\end{figure}

Cables are connected to the PMT bases from the back, as shown in figure~\ref{fig:facility_cabling}.
Wiring harnesses are attached to the back of the mounting rails with 3D-printed clamps, avoiding mis-cabling.
Each PMT is identified by a barcode.
The positions on the rails, the rails themselves, and the positions within the rack are also equipped with barcode identifiers.
We associate PMTs with positions and rails with rack positions whenever a PMT or a rail is moved.
During the measurement campaign, this system ensures reliable monitoring of the association between the PMT and the DAQ channel.

\subsection{Data Acquisition}

The data acquisition is realized by four mDOM mainboards that can each support up to \num{24} PMT channels~\cite{mDOM_mainboard_ICRC2023}.
Using generic mDOM functionality within the setup reduces cost and complexity and improves the comparability of these tests with the final operation within integrated mDOMs.
As a side-effect, the firmware development and mDOM integration have greatly benefited from this extensive use of the specific mDOM DAQ.

Within each mDOM mainboard, the signal is fed through a pulse shaping and amplification circuit and then digitized by a 12-bit ADC at \SI{120}{\mega\hertz}~\cite{mDOM_mainboard_ICRC2023}.
To improve timing capabilities, an adjustable discriminator on each channel is sampled at \SI{960}{\mega\hertz}~\cite{mDOM_mainboard_ICRC2023}.
One channel on each mainboard is used to trigger on pulses provided by the light source system, enabling a synchronization of all four boards with the pulsed light emission.
The mainboards are connected to a so-called mini-FieldHub~\cite{mDOM_Testing_ICRC2023}, an interface between the mainboards and the DAQ PC.

\subsection{Light Source System}

\begin{figure}[htbp]
    \centering
    \includegraphics[width=.8\textwidth]{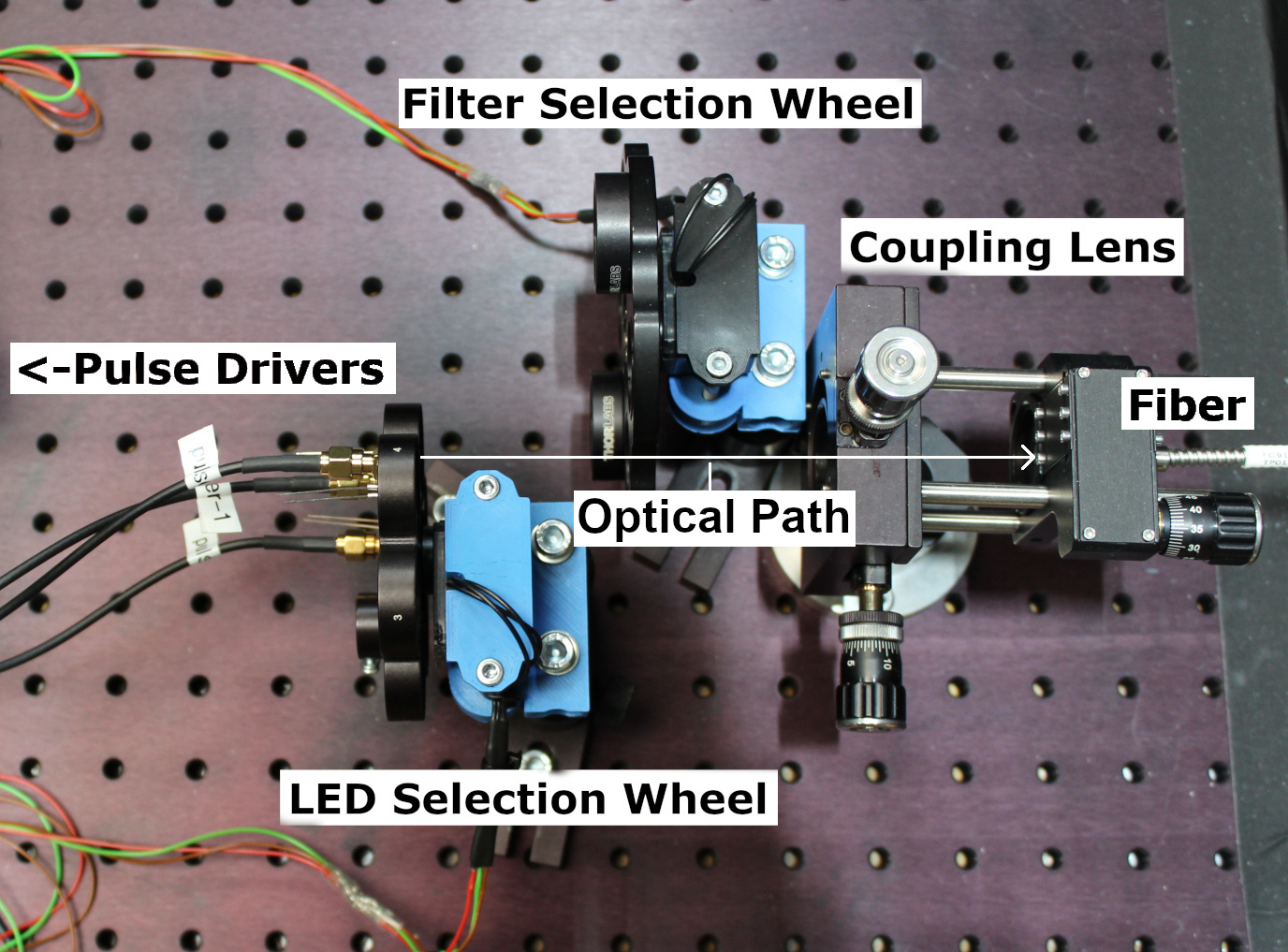}
    \caption{The Light Source System at the PMT Testing Facility. LEDs and optional neutral density filters are mounted to a selection wheel steered by stepper motors. The filters can be used to change the pulse intensity. The emitted light is coupled into an optical fiber.} 
    \label{fig:facility_light_source_system}
\end{figure}

The light source system consists of nanosecond pulse drivers~\cite{Rongen_2018}, an LED selection system, an optical fiber, and a diffuser.
The pulse drivers are controlled by the DAQ PC and drive two LEDs of the same type at wavelengths $\lambda_1=\SI{375}{\nano\meter}$~\cite{Roithner_XSL-375-3E} and one LED at $\lambda_2=\SI{505}{\nano\meter}$~\cite{Roithner_B3B-443-B505}; they also produce a time reference signal synchronous to the light emission.
The time jitter of the LED emission is $\delta_\mathrm{t,\,LED}\approx\SI{1}{\nano\second}$.
One pulse driver is tuned to produce very bright pulses in one of the LEDs at $\lambda_1=\SI{375}{\nano\meter}$~\cite{Roithner_XSL-375-3E}; the other pulse drivers are configured to provide dim brightnesses resulting in Single Photoelectron (SPE) levels at the PMTs.
Figure~\ref{fig:facility_light_source_system} shows the LED selection system: The LEDs are mounted to a selection wheel controlled by a stepper motor.
The produced light is coupled into an optical fiber using a fiber launch assembly.
The light is guided through this fiber to a \SI{48}{\mm} diameter PTFE diffuser adapted from the Precision Optical Calibration Module~\cite{POCAM_paper}, seen on the left of figure~\ref{fig:facility_pmt_rack}.
The symmetry axis of the diffuser defines the optical axis, which coincides with the closest distance to the PMT rack.

\subsection{Commissioning}

The amount of light received at each position of the PMT rack had to be calibrated to allow for measurements of PMT photodetection efficiencies.
The distance of the PMT to the integrating sphere, the angular emission profile of the integrating sphere, and the inclination angle of the PMTs to the received light front introduce a position dependence to the received brightness.
To allow correcting for this, we have calibrated the dependence of the relative brightness on the distance $d$ from a PMT to the intersection of the optical axis with the rack plane, i.\,e., the point nearest the integrating sphere.
The parameterization is given by equation~\ref{eq:correction_factor}, where $R_0=\SI{1.37}{\meter}$ is the distance between the diffuser and PMT rack and generic parameters $a$ and $c$.

\begin{equation}
    f_\mathrm{corr}(d) = a \cdot \left(\frac{R_0}{\sqrt{R_0^2 + d^2}}\right)^c
    \label{eq:correction_factor}
\end{equation}

\begin{figure}[htbp]
    \centering
    \includegraphics[width=.9\textwidth]{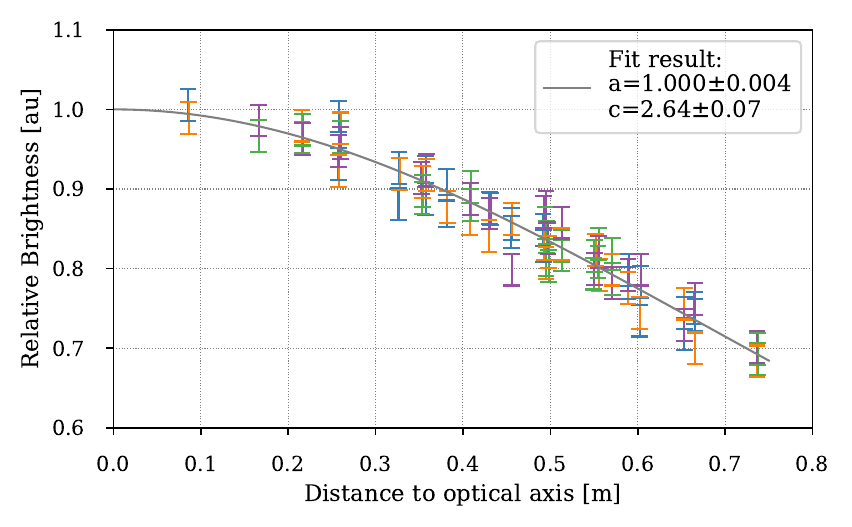}
    \caption{Flat-field illumination correction of the PMT Rack. Each color represents a PMT used in the calibration; the errorbars only contain statistical errors. The grey line shows a fit of equation~\ref{eq:correction_factor} to the data.}
    \label{fig:facility_flat_field_correction}
\end{figure}

The relative illuminations were measured with six PMTs in repeated measurements at different positions and at a constant illumination level.
The received brightness was inferred from the number of pulses measured in each PMT.
Two PMTs remained at the center of the rack, and the other four were moved through the rack.
The PMTs at the center were used as a reference and allowed correcting for possible brightness variations of the light source system.
Figure~\ref{fig:facility_flat_field_correction} shows the measured relative brightnesses as a function of the distance to the optical axis decreasing by up to \SI{30}{\percent} towards the edges.
Equation~\ref{eq:correction_factor} was fitted to the data and is used to compute individual correction factors for illumination at each position in the rack; the uncertainty of these correction factors is below \SI{2}{\percent}.

To convert the measured relative brightness to a photodetection efficiency, absolutely calibrated reference PMTs are required.
The manufacturer provided absolute cathode-current measurements of the full-face quantum efficiencies for \SI{10}{\percent} of the PMTs.
We have selected reference PMTs based on two criteria: Long-term PMT gain and photodetection efficiency stability~\cite{joelle-thesis}.
In every testing cycle, four reference PMTs are used in the PMT rack in randomly selected positions.
The reference PMTs were exchanged between the testing facilities to cross-calibrate the amplification gain and dark rate measurements.

Variations in the amplification gain of the analog frontend channels of the mDOM mainboards~\cite{mDOM_mainboard_ICRC2023} were calibrated using twelve PMTs moving through the PMT rack, connecting the same PMT to multiple channels.
The supply voltages of the PMTs were kept constant, and the PMT amplification gain was measured with SPE spectra.
The measured amplification gains were normalized to a channel in the PMT rack, and the absolute scale was calibrated with measurements using a commercial FADC module~\cite{CAEN_v1730}.
This calibration reduces the systematic uncertainty in the measurements of the PMT amplification gains from \SI{6}{\percent} to less than \SI{2}{\percent}.

    \section{Measurement Procedures}\label{sec:measurements}

\begin{figure}[htbp]
    \centering
    \includegraphics[width=.9\textwidth]{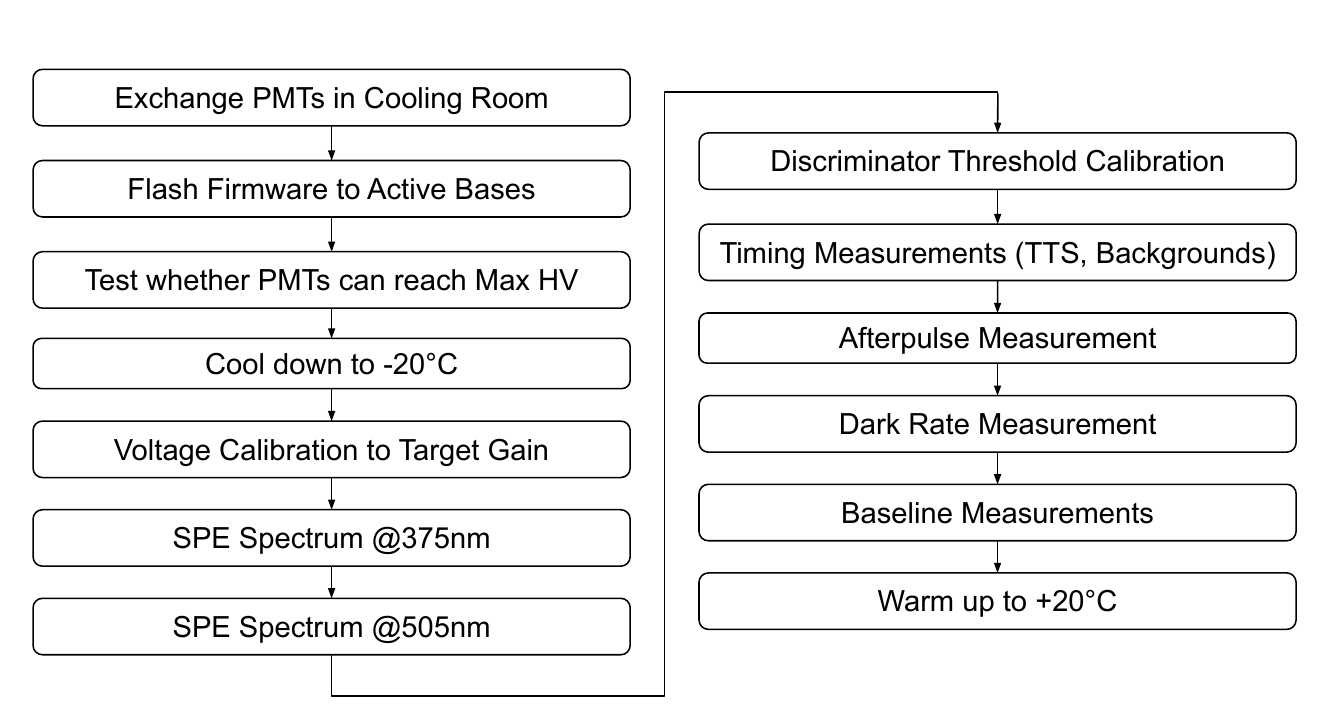}
    \caption{Structure of a full PMT testing cycle.}
    \label{fig:testing_cycle_structure}
\end{figure}

All functions of the testing facility can be controlled by software, allowing for fully automated measurement procedures.
The standard procedure of a single measurement cycle is shown in figure~\ref{fig:testing_cycle_structure}.
A cycle starts with removing the PMTs from the previous testing cycle from the cooling room and installing the prepared mounting rails on which the new PMTs are mounted.
After the PMTs are connected to the cables and the door of the cooling room is closed, the software takes over: Firmware is flashed to the microcontroller on the PMT base, and the maximum specified high voltage is applied to the PMTs.
The calibration measurements are performed once the facility is cooled down to \SI{-20}{\degreeCelsius}.
When all measurements have finished, the high voltage is turned off, and the facility is warmed to room temperature.
A testing cycle typically lasts just less than \SI{24}{\hour}.

The software design follows the structure of the testing cycle: A linear set of tasks is performed, ranging from data-taking and analysis of that data to controlling the facility's temperature and devices.
Results are stored in a database.
In the following, we describe the data recorded at the facility and the associated analyses.

\begin{figure}[htbp]
    \centering
    \includegraphics[width=.9\textwidth]{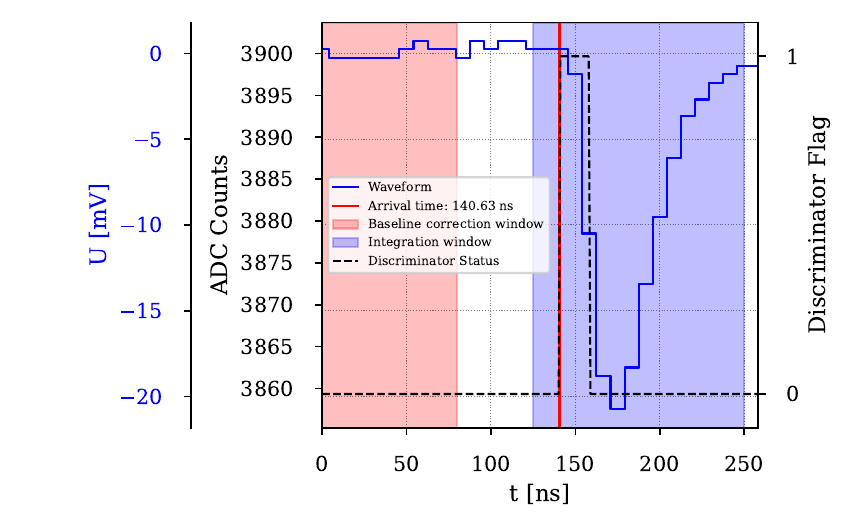}
    \caption{Example recorded waveform of a tested PMT. The blue line is the recorded waveform; the red area is used for estimating the baseline level; the blue area is the charge integration range; the black line is the status of the discriminator output; and the red line is the estimated pulse arrival time. The PMT signal is shaped before digitization on the mDOM mainboard.}
    \label{fig:waveform}
\end{figure}

In standard operation, the PMTs are illuminated by the light source system, and the digitization of the PMT signals is triggered by a pulse synchronized to the pulsed light emission.
The generically recorded data structure is a PMT waveform.
Figure~\ref{fig:waveform} shows an example waveform with ADC counts calibrated to voltages and the readout of the discriminator output.
The baseline level is estimated from the same recorded waveform as the median voltage measured in a time window of \SI{80}{ns} width, indicated by the red-shaded region, before the expected arrival of light signals.
After correction of the baseline level, the charge of the PMT pulse is extracted by integrating over the blue-shaded region of \SI{125}{\nano\second} width.
The length of the integration window has been chosen such that it includes the entire signal, including arrival time variations.
With the known input impedance of the amplifier, the integrated voltage is converted to a charge.

\subsection[SPE Spectra]{Single-Photoelectron Spectra}

\begin{figure}[htbp]
    \centering
    \includegraphics[width=.8\textwidth]{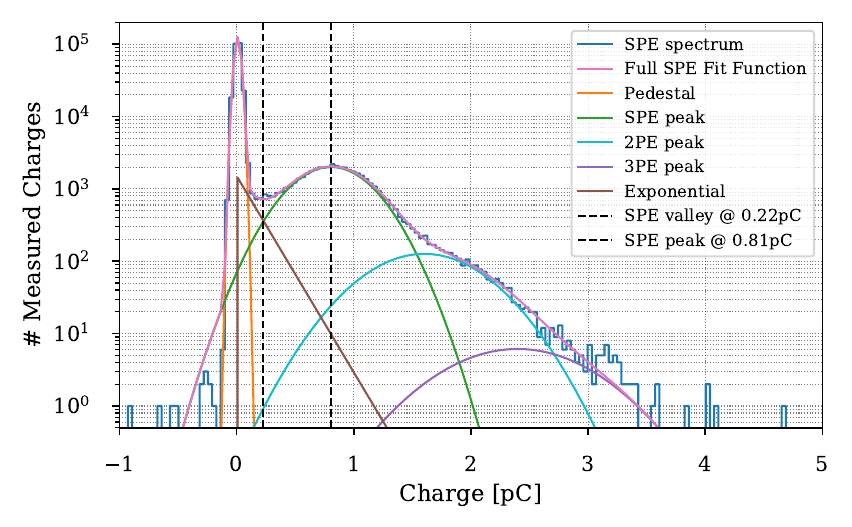}
    \caption{An example SPE spectrum of an mDOM PMT is shown in blue. The fit function in equation~\ref{eq:spe_fit_function} is shown in colored lines. The vertical, dashed lines indicate the valley and SPE peak.}
    \label{fig:spe_spectrum}
\end{figure}

We measure Single-Photoelectron (SPE) spectra by adjusting the brightness of the light source system such that roughly \SI{10}{\percent} of the recorded waveforms contain a PMT pulse.
The digitization of waveforms is triggered by the synchronization pulse from the light source system; the pulse rate is set to \SI{5}{\kilo\hertz}.
The integrated charges of the waveforms are histogrammed.
Figure~\ref{fig:spe_spectrum} shows an example histogram of the charges from \num{300000} recorded waveforms.
The pedestal peak at $q_\mathrm{ped}=\SI{0}{\pico\coulomb}$ consists of waveforms without a PMT pulse, and the peak at $q_\mathrm{SPE}=\SI{1}{\pe}\approx\SI{0.8}{\pico\coulomb}$ is comprised of single photon detections.

Equation~\ref{eq:spe_fit_function} is fitted to the histogram for extracting SPE properties~\cite{BELLAMY1994468}.
It consists of two templates for differently modeled signal and noise components.
The first component represents the pedestal peak, the SPE peak, and peaks of higher order corresponding to multiple photoelectrons.
It is represented by a convolution of a Poissonian with normalization $A$, the mean number of detected photons per waveform $\mu$, and a series of Gaussian distributions.
The Gaussians are described by the width $\sigma_0$ and position $q_0$ of the pedestal and the width $\sigma_1$ and position $q_\mathrm{SPE}$ of the SPE peak.
The positions and widths of higher-order terms are scaled from these parameters.
Including the pedestal peak in the fit gives the level of electronic noise in the waveform baselines, which is accounted for in the widths of the following Gaussians, and helps us to constrain the measured $\mu$.
The second component describes poorly amplified photoelectrons by a zero-cutoff exponential with normalization $B$ and shape $\tau$ as free parameters.
It is not included in the higher-order terms; the effect on the measured gains and on $\mu$ is less than \SI{2}{\percent}.

\begin{align}
    f\left(q\right) = &A \cdot \sum_{n=0}^4 \frac{e^{-\mu} \mu^n}{n!} \frac{1}{\sqrt{2\pi\left(\sigma_0^2 + n\cdot\sigma_\mathrm{SPE}^2\right)}}\exp{\left(-\frac{\left(q - q_0 - n \cdot q_\mathrm{SPE}\right)^2}{2\left(\sigma_0^2 + n \cdot \sigma_\mathrm{SPE}^2\right)}\right)}\enspace+\label{eq:spe_fit_function}\\
    & B \cdot \exp{\left(-\frac{q}{\tau}\right)} \cdot \Theta\left(q\right)\nonumber
\end{align}

The gain $g=\frac{q_\mathrm{SPE}}{e}$ can be directly inferred from the SPE peak position and the elementary charge $e$.
The SPE charge resolution $\sigma_\mathrm{SPE} = \frac{\sigma_1}{q_\mathrm{SPE}}$ is computed as the intrinsic width of the SPE peak normalized to its position.
This normalization makes the SPE charge resolution more independent of the chosen PMT amplification gain.
The peak-to-valley ratio is calculated from the full function with the best-fit parameters.
The peak-to-valley ratio and SPE charge resolution recorded at the nominal gain are compared to the requirements in table~\ref{tab:specs}.

\subsection{Voltage Calibration}

The gain of the PMT is a function of the high voltage applied to the PMT and the material properties of the dynodes~\cite{Hamamatsu_PMT_Handbook}.
It typically follows a power law as in equation~\ref{eq:gain_power_law}.

\begin{equation}
    g = c \cdot U_\mathrm{dy10}^{k}
    \label{eq:gain_power_law}
\end{equation}

Here, $g$ is the PMT amplification gain, $c$ is a constant, $U_\mathrm{dy10}$ is the applied high voltage, and $k$ depends on the dynode material and typically is between 0.7 and 0.8~\cite{Hamamatsu_PMT_Handbook}.

\begin{figure}[htbp]
    \centering
    \includegraphics[width=.9\textwidth]{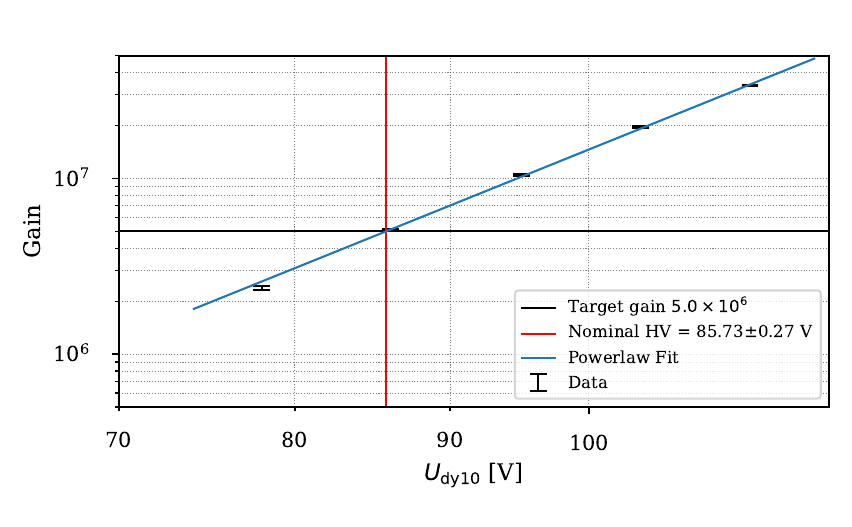}
    \caption{Example gain calibration curve for an mDOM PMT. The black points are measured gains; the blue line is a fit of equation~\ref{eq:gain_power_law} to the data; the horizontal black line indicates the nominal gain of $g_\mathrm{nom}=\SI{5e6}{\relax}$; the vertical red line indicates the extracted nominal voltage $U_\mathrm{dy10\,nom}$.}
    \label{fig:nominal_voltage_calibration}
\end{figure}

Five SPE spectra at high voltages covering the whole specification range are recorded to calibrate the PMT gain.
The gain is extracted for each SPE spectrum, and equation~\ref{eq:gain_power_law} is fitted to them as a function of the applied high voltage $U_\mathrm{dy10}$.
Figure~\ref{fig:nominal_voltage_calibration} shows an example of such a gain calibration.
The nominal voltage $U_\mathrm{dy10,\,nom}$ is the voltage at which the PMT reaches the nominal gain of $g_\mathrm{nom}=\SI{5e6}{\relax}$.
It is extracted from the power law fit and shown as a red vertical line in figure~\ref{fig:nominal_voltage_calibration}.
All following measurements are made with each PMT operating at its individual nominal voltage.

\subsection{Photodetection Efficiency}

The measurement of the quantum efficiency of the PMTs is achieved indirectly: Four previously selected reference PMTs with calibrated quantum efficiencies provided by the manufacturer are mounted in the PMT rack for every calibration cycle.
The observed fraction of photons detected by each PMT is proportional to quantum efficiency, collection efficiency, and the effective cathode area.
Assuming variations in collection efficiencies and effective cathode areas to be small across the total ensemble of PMTs, we compare detection fractions with those of the reference PMTs.
The mean number of detected photoelectrons per waveform $\mu$ inferred from the simultaneously measured SPE spectra, see equation~\ref{eq:spe_fit_function}, are corrected for the channel-specific light yield $f_\mathrm{corr}\left(d\right)$ shown in figure~\ref{fig:facility_flat_field_correction}.
Then, the measured quantum efficiency is given by

\begin{equation}
    QE_\mathrm{PMT} = QE_\mathrm{ref.\,PMT} \frac{\mu_\mathrm{PMT}}{f_\mathrm{corr,\,PMT}} \frac{f_\mathrm{corr,\,ref.\,PMT}}{\mu_\mathrm{ref.\,PMT}}.
    \label{eq:computation_quantum_efficiency}
\end{equation}

The median of the measured quantum efficiencies relative to all reference PMTs is compared to the requirements in table~\ref{tab:specs}. This measurement is performed at the wavelengths $\lambda_1=\SI{375}{\nano\meter}$ and $\lambda_2=\SI{505}{\nano\meter}$.

\subsection{Discriminator Calibration}

\begin{figure}[htbp]
    \centering
    \subfigure[Computation of the charge threshold with SPE spectra recorded at a gain of $g=\SI{5e6}{\relax}$. The blue histogram contains all recorded waveforms; the orange histogram contains only waveforms with a discriminator crossing. The red line is the ratio of the histograms; the vertical grey line is the extracted charge threshold; the dashed horizontal line indicates a ratio of \SI{0.5}{\relax}.]{
        \includegraphics[width=0.8\textwidth]{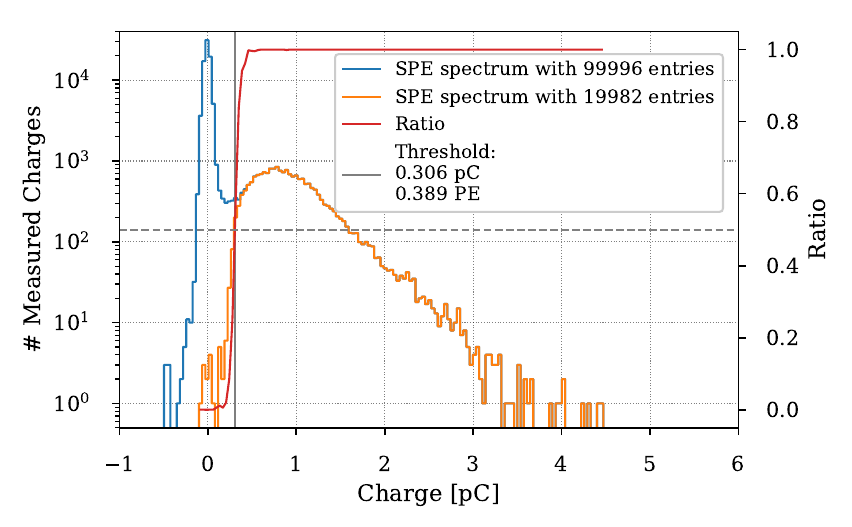}
    }
    \subfigure[Calibration of the DAC input setting to a charge threshold. The blue points are measured charge thresholds for five DAC settings; the orange line is a linear fit to the data; the vertical green line is the target charge threshold of \SI{0.5}{\pe}; the dashed horizontal line indicates the extracted DAC setting.]{
        \includegraphics[width=0.8\textwidth]{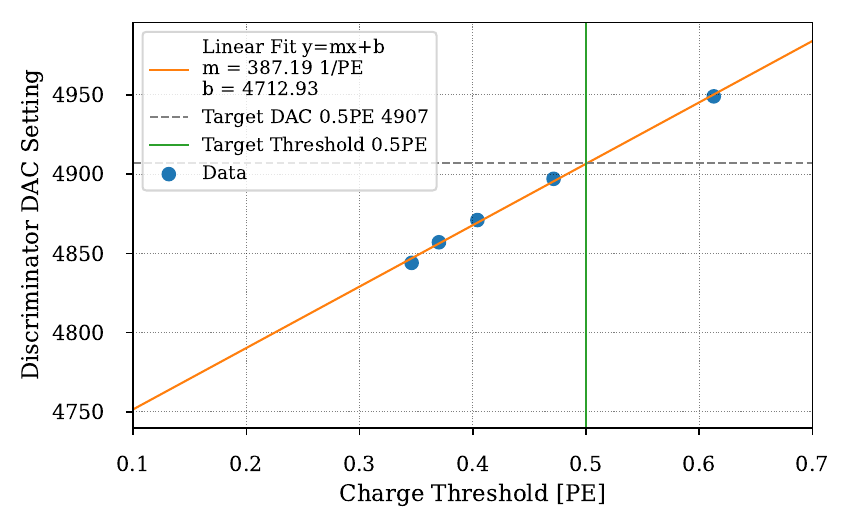}
    }
    \caption{Example calibration of an mDOM mainboard discriminator.}
    \label{fig:discriminator_calibration}
\end{figure}

Each channel on the mDOM mainboard has a discriminator in the analog frontend~\cite{mDOM_mainboard_ICRC2023}.
The PMT signal pulse is compared to a reference voltage generated by a DAC.
The input of the reference DAC must be calibrated to a charge threshold.
After the calibration of the PMT voltage, five SPE spectra with varying DAC input settings are recorded.
The top of figure~\ref{fig:discriminator_calibration} illustrates the extraction of the charge threshold from the ratio of two histograms: One with all waveforms and one with only those waveforms with a discriminator crossing.
The charge at which the discriminator is crossed \SI{50}{\percent} of the time is converted to units of photoelectrons.

The DAC input settings are fitted with a linear function as a function of the measured charge thresholds, as shown at the bottom of figure~\ref{fig:discriminator_calibration}.
The target DAC setting can easily be inferred from the best-fit function.
We calibrate the discriminators to a charge threshold of \SI{0.5}{\pe} instead of \SI{0.2}{\pe} as specified in table~\ref{tab:specs} to avoid triggering on electronic noise introduced by the laboratory environment.

\subsection{Timing and Backgrounds}

\begin{figure}[htbp]
    \centering
    \includegraphics[width=0.9\textwidth]{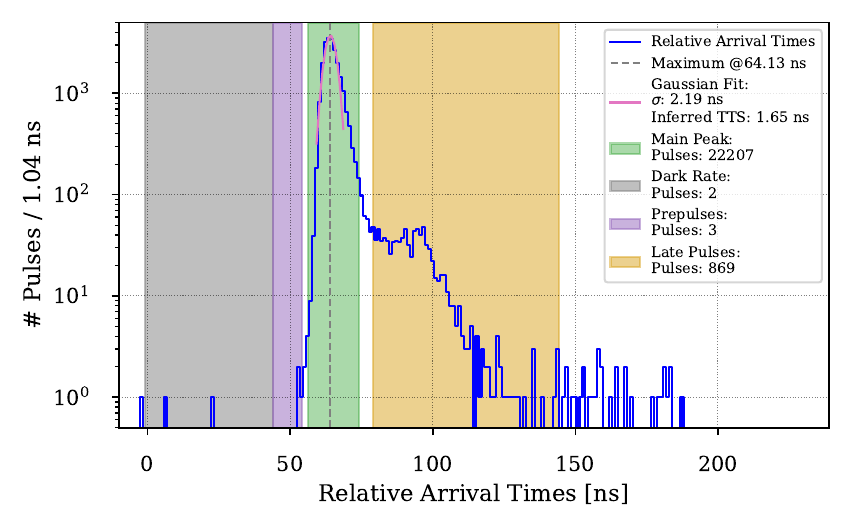}
    \caption{An example mDOM PMT arrival time distribution, relative to the synchronization signal, is shown in blue. The pink line indicates a Gaussian fit to the peak, which is used to estimate the TTS. The green region is used to count the number of standard SPE pulses; the purple region is used to count prepulses; the beige region is used to count late pulses; the grey region is used to estimate the contribution from dark rate pulses.}
    \label{fig:arrival_time_spectrum}
\end{figure}

The transit-time spread (TTS) is measured using arrival times of PMT pulses relative to the emission time of the photons from the light source system.
One channel on each mDOM mainboard receives and digitizes synchronization pulses serving as the time reference.
The source brightness is adjusted to the SPE level.
The signal arrival time is taken from the discriminator, which was previously calibrated to a threshold of \SI{0.5}{\pe}.
In the case of multiple pulses within the waveform, all arrival times are extracted.
The PMT arrival times are corrected for the arrival times of the synchronization pulses.
Figure~\ref{fig:arrival_time_spectrum} shows an example histogram of the corrected arrival times.
A Gaussian is fitted to the main peak, and the width $\sigma$ is extracted.
We estimate the influence on the time jitter by the DAQ sampling resolution, the integrating sphere, and variations in signal amplitudes to $\delta_\mathrm{t,\,other}\approx\SI{1}{\nano\second}$.
We quadratically subtract these effects and the time jitter of the LED emission ($\delta_\mathrm{t,\,LED}\approx\SI{1}{\nano\second}$) from $\sigma$ to compute the transit time spread.
The main peak position (typically at \SI{65}{\nano\second}) reflects both the mean transit time and the sum of other propagation delays in the measurement setup.

The arrival time histogram is also used to estimate the prepulse and late pulse fractions.
The purple band in figure~\ref{fig:arrival_time_spectrum} indicates the prepulse region: It is defined as a window [\SI{-20}{\nano\second},\,\SI{-10}{\nano\second}] relative to the main peak maximum; the late pulse region [\SI{15}{\nano\second},\,\SI{80}{\nano\second}] relative to the main peak is shown in beige.
The corresponding fractions are computed as the ratio of the number of pulses in those regions relative to the number of pulses arriving in the main transit time distribution marked in green.
The contamination from dark rate pulses is estimated in the grey region left of the main peak.
The afterpulse fraction is measured similarly with longer waveform readout windows of $\SI{12}{\micro\second}$ and is not shown here.

The dark pulse rates are measured by counting discriminator threshold crossings on each channel.
The discriminator thresholds are calibrated to an equivalent of \SI{0.5}{\pe} to avoid contributions from electronic noise introduced by the laboratory environment.
However, the specification for the dark rate requires a lower threshold of \SI{0.2}{\pe}.
Therefore, we apply PMT-individual correction factors, which are computed as the ratio of integrals over the SPE peaks in the previously recorded SPE spectra at the nominal voltage starting at \SI{0.2}{\pe} and \SI{0.5}{\pe}.

\begin{equation}
    R_\mathrm{corr} = \frac{\int_{\SI{0.2}{\pe}}^\infty{}f\left(q\right)dq}{\int_{\SI{0.5}{\pe}}^\infty{}f\left(q\right)dq} \approx 1.15
    \label{eq:dark_rate_correction_factor}
\end{equation}

The number of pulses is counted during 100 measurements of \SI{10}{\second} duration each.
Here, pulses occurring less than \SI{100}{\nano\second} after a count are not also counted (artificial deadtime).
The median of these count rates is compared to the requirement in table~\ref{tab:specs}.

    \section{Results}\label{sec:results}

The above-described measurement program has been performed for all \num{10427} PMTs.
In the following, we present selected results and discuss the compliance with requirements in table~\ref{tab:specs}.

\begin{figure}[htbp]
    \centering
    \includegraphics[width=.8\textwidth]{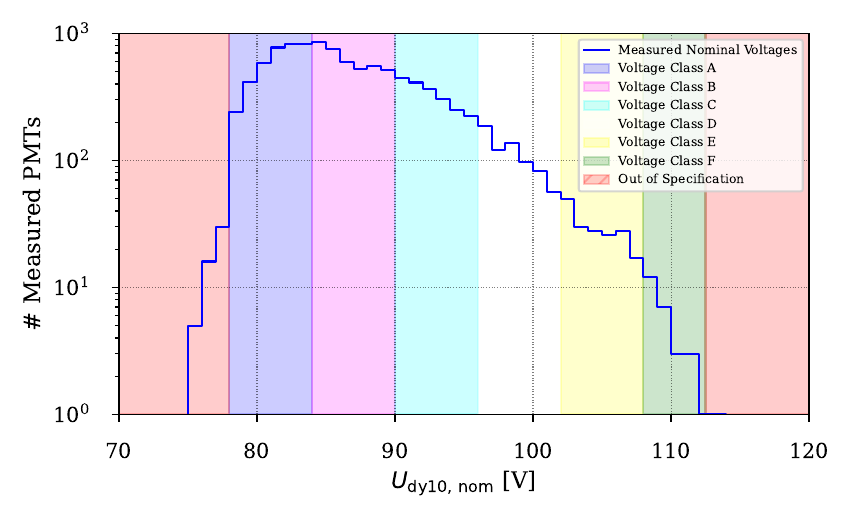}
    \caption{Histogram of measured nominal voltages for all PMTs for a gain of $g=\SI{5.0e6}{\relax}$. The red bands indicate the specification region; the other colored bands show the classes the PMTs are sorted into.}
    \label{fig:nominal_voltage_distribution}
\end{figure}

Figure~\ref{fig:nominal_voltage_distribution} shows the distribution of measured nominal voltages $U_\mathrm{dy10,\,nom}$.
The distribution smoothly covers the specified range.
The red shaded areas indicate voltages that are outside the specification range.
The colored bands show the voltage classes into which the PMTs are sorted.
The small number of PMTs in the distribution's tails outside the specification range are still accepted since no operational problems were observed.

\begin{figure}[htbp]
    \centering
    \includegraphics[width=.8\textwidth]{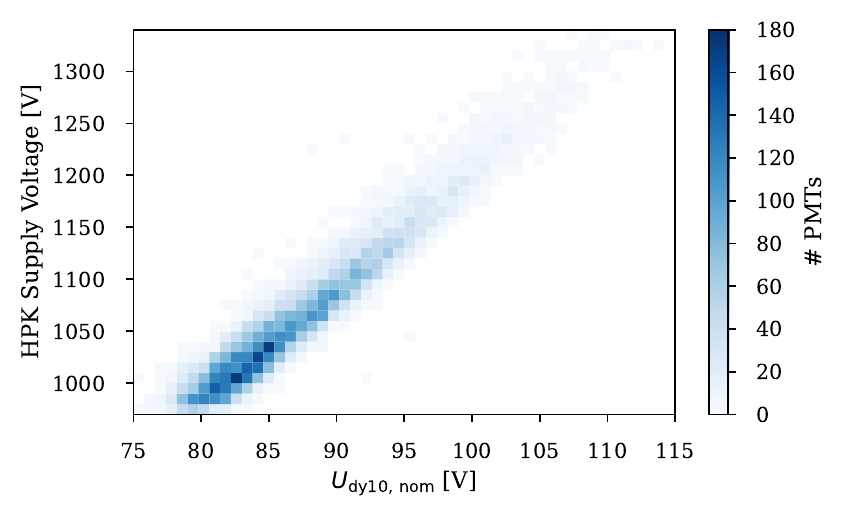}
    \caption{Correlation of measured nominal voltages with cathode voltage values reported by the manufacturer (HPK).}
    \label{fig:nominal_voltage_correlation}
\end{figure}

The manufacturer provides nominal supply voltages for the entire dynode system.
Figure~\ref{fig:nominal_voltage_correlation} shows a two-dimensional histogram of manufacturer-quoted and measured nominal voltage $U_\mathrm{dy10,\,nom}$.
The determined nominal voltages correlate strongly with the manufacturer values, indicating a good agreement between both measurements.

\begin{figure}[htbp]
    \centering
    \subfigure{
        \includegraphics[width=0.8\textwidth]{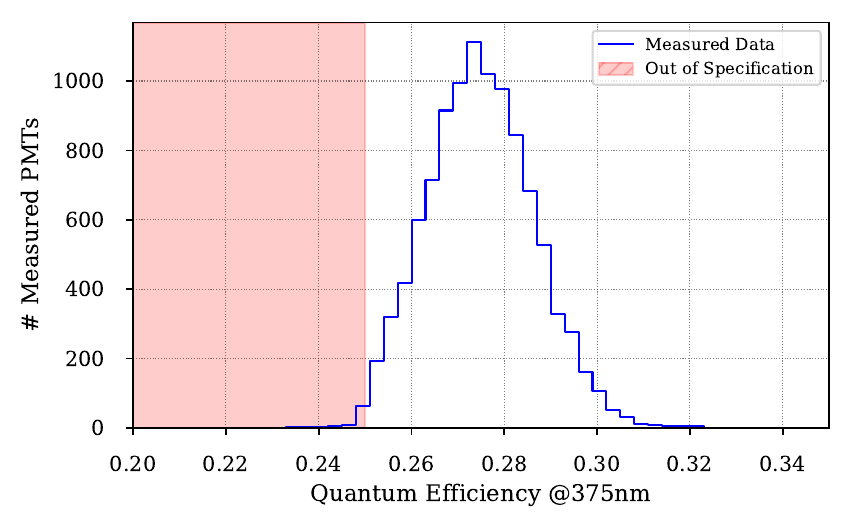}
    }
    \subfigure{
        \includegraphics[width=0.8\textwidth]{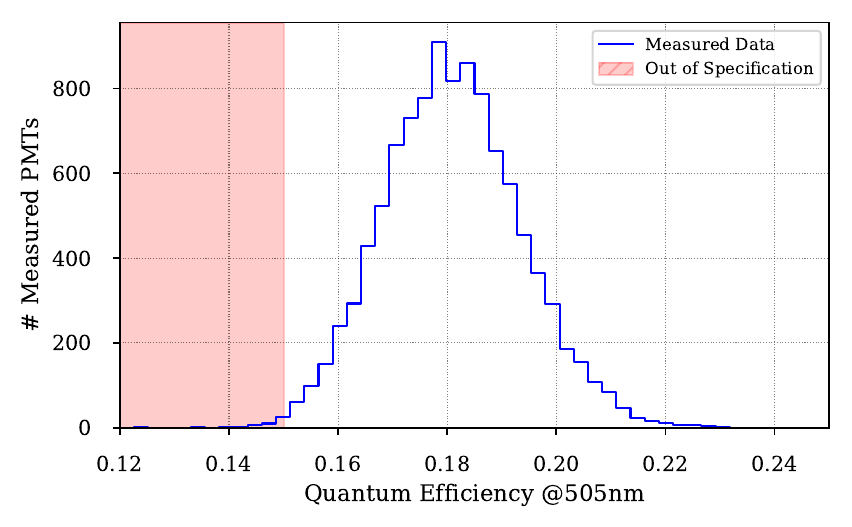}
    }
    \caption{Distributions of measured quantum efficiencies at $\lambda=375\,\mathrm{nm}$ (top) and $\lambda=505\,\mathrm{nm}$ (bottom).}
    \label{fig:results_quantum_efficiencies}
\end{figure}

Figure~\ref{fig:results_quantum_efficiencies} shows the distributions of the quantum efficiencies measured at $\lambda_1=375\,\mathrm{nm}$ and $\lambda_2=505\,\mathrm{nm}$.
The distributions are smooth, and less than \num{100} PMTs fail to meet the requirements in table~\ref{tab:specs}.
The distribution at $\lambda_1=375\,\mathrm{nm}$ has a mean of \SI{27.5}{\percent} and a standard deviation of \SI{1.1}{\percent}, the distribution at $\lambda_2=505\,\mathrm{nm}$  has a mean of \SI{18.1}{\percent} and a standard deviation of \SI{1.3}{\percent}.

\begin{figure}[htbp]
    \centering
    \subfigure{
        \includegraphics[width=0.8\textwidth]{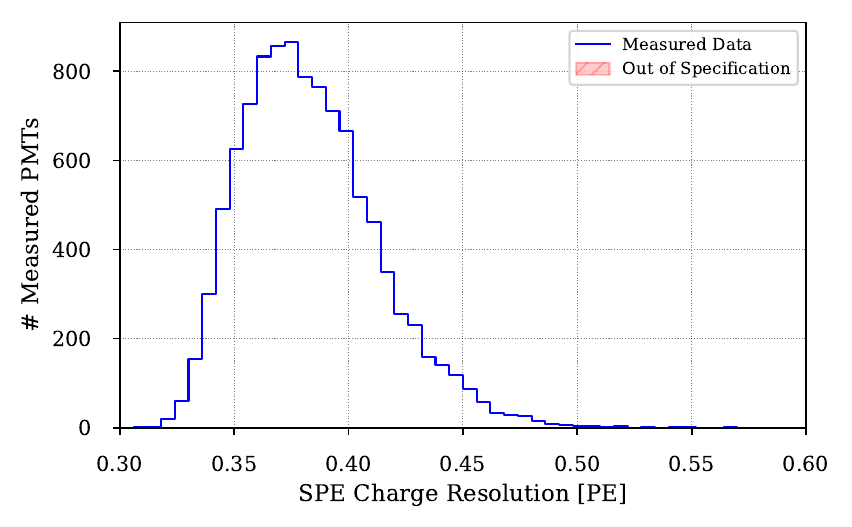}
    }
    \subfigure{
        \includegraphics[width=0.8\textwidth]{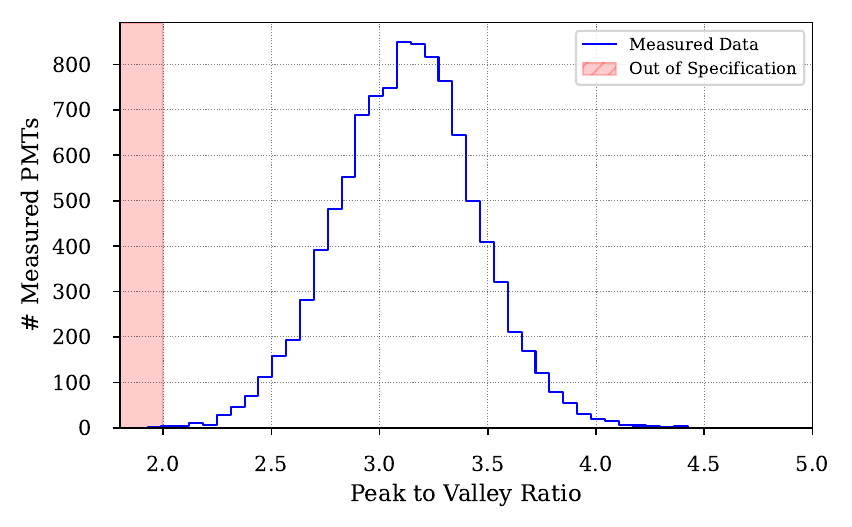}
    }
    \caption{Distributions of measured SPE charge resolutions (top) and peak-to-valley ratios (bottom).}
    \label{fig:results_spe_properties}
\end{figure}

Figure~\ref{fig:results_spe_properties} shows the distributions of measured SPE charge resolutions and peak-to-valley ratios.
Both distributions are smooth, and less than \num{50} PMTs fail to meet the requirements in table~\ref{tab:specs}.

\begin{figure}[htbp]
    \centering
    \subfigure{
        \includegraphics[width=0.8\textwidth]{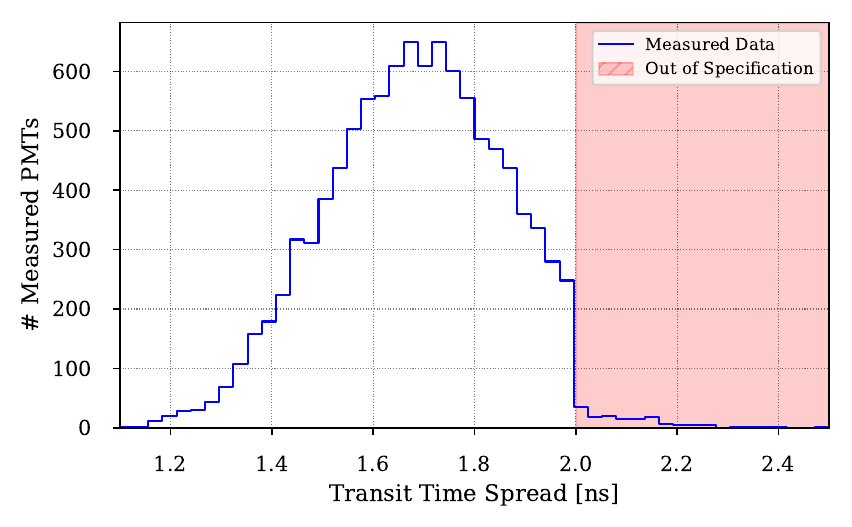}
    }
    \subfigure{
        \includegraphics[width=0.8\textwidth]{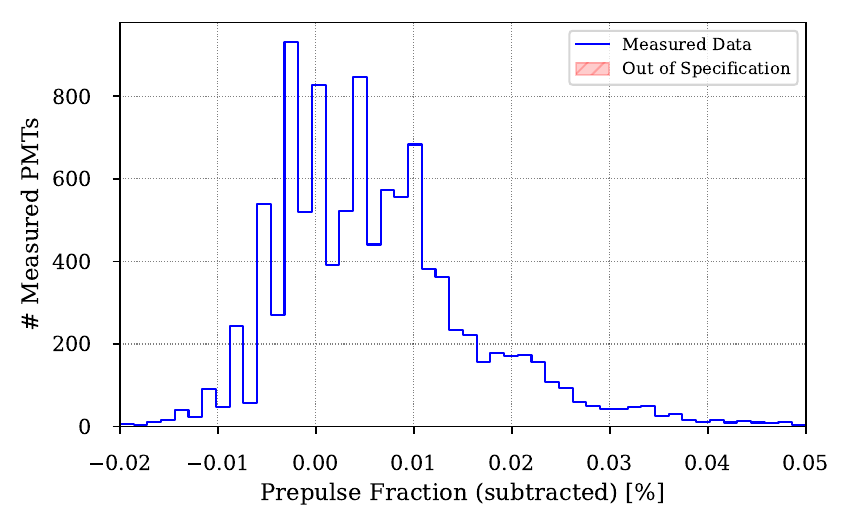}
    }
    \caption{Distributions of measured transit time spreads (top) and prepulse fractions (bottom). The edge in the transit time spread distribution results from multiple testing of failed PMTs with limited measurement precision.}
    \label{fig:results_timing_characteristics_1}
\end{figure}

\begin{figure}[htbp]
    \centering
    \subfigure{
        \includegraphics[width=0.8\textwidth]{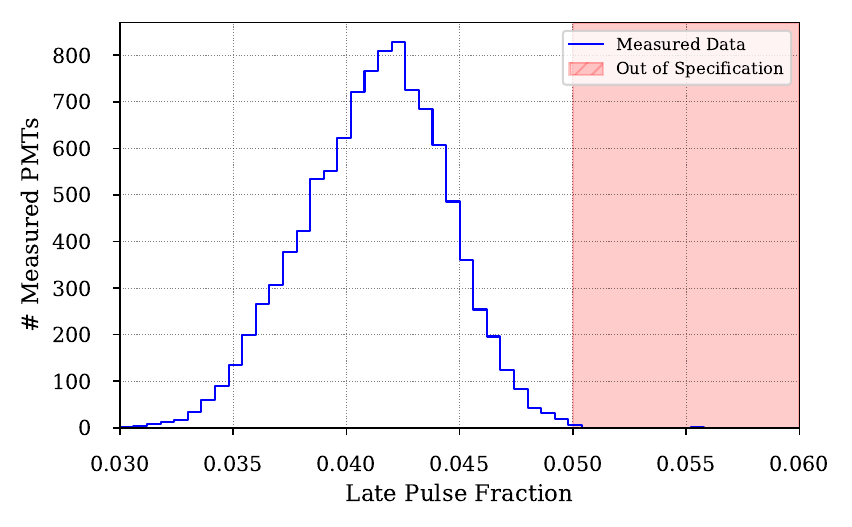}
    }
    \subfigure{
        \includegraphics[width=0.8\textwidth]{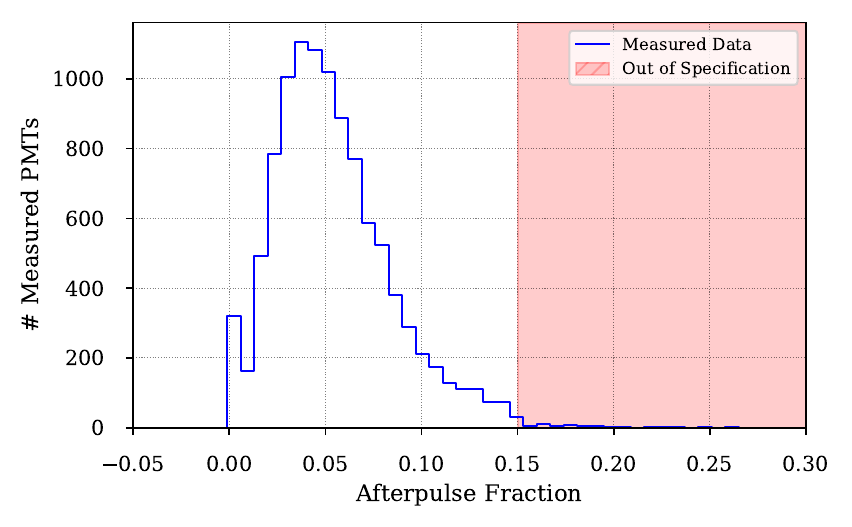}
    }
    \caption{Distributions of measured late pulse (top) and afterpulse fractions (bottom).}
    \label{fig:results_timing_characteristics_2}
\end{figure}

Figures~\ref{fig:results_timing_characteristics_1} and~\ref{fig:results_timing_characteristics_2} show the distributions of measured timing and signal-correlated background performance parameters.
The measured transit time spreads peak at $1.7\,\mathrm{ns}$ and have a clear edge at the specification boundary.
This results from the limited measurement resolution of $\sigma_\mathrm{TTS}=250\,\mathrm{ps}$ for the transit time spread.
Retesting failing PMTs can lead to a smaller measured transit time spread complying with the specification; PMTs complying with the specifications are not tested again.

The measured prepulse, late pulse, and afterpulse fractions are shown in figures~\ref{fig:results_timing_characteristics_1} and~\ref{fig:results_timing_characteristics_2}.
Negative values in the prepulse fractions result from statistical fluctuations in the dark rate corrections in the case of very few measured prepulses which also results in the spikes in the distribution.
The measured prepulse fractions are well below \SI{1}{\percent}, the late pulse fraction distribution has a mean of \SI{4.1}{\percent} with a standard deviation of \SI{0.3}{\percent}, and the afterpulse fraction distribution has a mean of \SI{5.4}{\percent} with a standard deviation of \SI{3.1}{\percent}.

\begin{figure}[htbp]
    \centering
    \includegraphics[width=.8\textwidth]{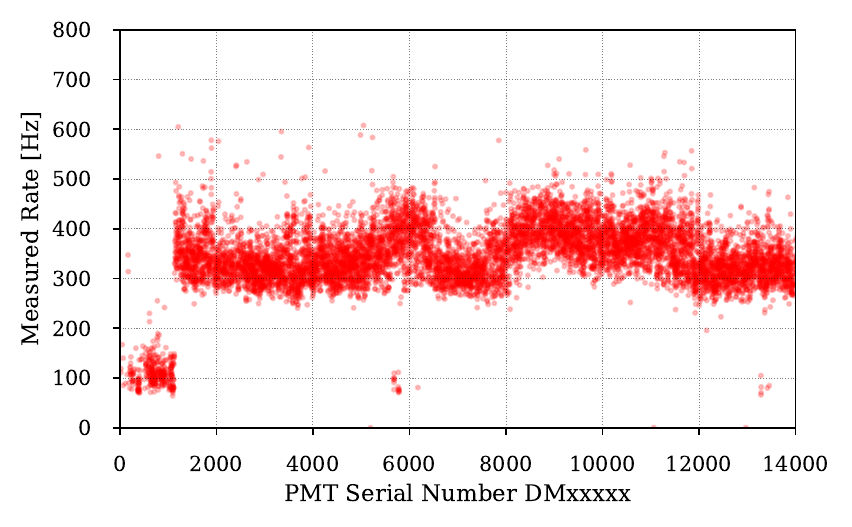}
    \caption{Measured dark rates of the mDOM PMTs as a function of PMT serial number. Most PMTs have higher noise rates than originally specified. This results from increased concentrations of radioactive material in the PMT glass.}
    \label{fig:darkrates}
\end{figure}

Figure~\ref{fig:darkrates} shows the results of the dark rate measurements as a function of the PMT serial number.
Most PMTs have dark rates higher than the requirements of $\leq\SI{150}{\hertz}$.
There is a clear break in the dark rates at the PMT serial number DM01130.
The increased dark rates result from increased concentrations of radioactive materials in the PMT glass~\cite{munland_phd}.
Based on initial measurements of fully assembled mDOMs, we expect a total dark rate per integrated PMT of $\sim\SI{750}{\hertz}$ in deployed modules. This includes radioactive decays in the pressure vessel of the mDOMs. The contributed fraction from the additional noise component amounts to \SI{25}{\percent}. This noise increase can be compensated with modifications to the in-module data compression and transfer strategies. Thus, these PMTs were deemed usable and were accepted.

Table~\ref{tab:result_statistics} shows the number of compliant PMTs for all tested specifications.
Except for the dark rates, all non-conforming PMTs are close to the specification boundaries and, therefore, have still been accepted.
However, these PMTs will be used in spare mDOMs.
We observe a slight enhancement of correlated failures: Of the non-compliant \num{314} PMTs, disregarding the failures in the dark rate, \num{73} PMTs fail to meet two or more specifications.
Five PMTs were found to have more severe communication and performance problems.
These PMTs have been sent back to the manufacturer.

\begin{table}[htbp]
    \centering
    \caption{Statistics of the PMT testing results. PMTs with failed measurements are included as non-compliant.}
    \begin{tabular}{ c || c | c | c }
        Specification & \# compliant PMTs & \# non-compliant PMTs & \# accepted PMTs \\\hline\hline
        High Voltage @ Gain \SI{5e6}{\relax}  & 10349 & 78 & 10423 \\
        Quantum Efficiency & 10336 & 91 & 10427 \\
        TTS ($\sigma$) & 10255 & 172 & 10427 \\
        SPE Charge Resolution ($\sigma_\mathrm{SPE}$) & 10396 & 31 & 10427 \\
        Peak-to-Valley-Ratio & 10390 & 37 & 10427 \\
        Prepulses & 10383 & 44 & 10427 \\
        Late Pulses & 10381 & 46 & 10427 \\
        Afterpulses & 10317 & 110 & 10427 \\
        \hline
        Subtotal & 10113 & 314 & 10423 \\
        \hline
        Dark Rate & 351 & 10076 & 10426 \\
        \hline\hline
        Total & 147 & 10280 & 10422
    \end{tabular}
    \label{tab:result_statistics}
\end{table}
    \FloatBarrier
    \section{Summary}\label{sec:summary}

We have presented the acceptance tests for the PMTs of the mDOMs of the IceCube Upgrade.
The central challenge of \num{10427} PMTs being tested on a short timescale has been met with a modular design of the PMT testing facility and full automation of the testing procedures.
The facilities are optimized to provide a high throughput of $\mathcal{O}\left(100\right)$ PMTs within one day.
Two facilities have been constructed, resulting in a throughput of \num{1000} PMTs per week.
With this, the primary testing campaign of all PMTs was accomplished within a few months.
We have discussed the measurement strategies for the PMT performance parameters and presented the results.
The recorded data were used to decide on the usability of the PMTs in mDOMs; five PMTs have been rejected and sent back to the manufacturer.

The testing facilities have been designed to be adaptable to other PMTs.
Two prototype PMTs for the IceCube-Gen2~\cite{IceCube_Gen2_white_paper} project, as well as \num{371} PMTs for the P-ONE~\cite{P-ONE_whitepaper} project, have successfully been tested at one of the testing facilities~\cite{halve_phd}.

    \acknowledgments
    The IceCube collaboration acknowledges the significant contributions to this manuscript from Lasse Halve and Johannes Werthebach.
We acknowledge the support from the following agencies:
USA {\textendash} U.S. National Science Foundation-Office of Polar Programs,
U.S. National Science Foundation-Physics Division,
U.S. National Science Foundation-EPSCoR,
U.S. National Science Foundation-Office of Advanced Cyberinfrastructure,
Wisconsin Alumni Research Foundation,
Center for High Throughput Computing (CHTC) at the University of Wisconsin{\textendash}Madison,
Open Science Grid (OSG),
Partnership to Advance Throughput Computing (PATh),
Advanced Cyberinfrastructure Coordination Ecosystem: Services {\&} Support (ACCESS),
Frontera computing project at the Texas Advanced Computing Center,
U.S. Department of Energy-National Energy Research Scientific Computing Center,
Particle astrophysics research computing center at the University of Maryland,
Institute for Cyber-Enabled Research at Michigan State University,
Astroparticle physics computational facility at Marquette University,
NVIDIA Corporation,
and Google Cloud Platform;
Belgium {\textendash} Funds for Scientific Research (FRS-FNRS and FWO),
FWO Odysseus and Big Science programmes,
and Belgian Federal Science Policy Office (Belspo);
Germany {\textendash} Bundesministerium f{\"u}r Bildung und Forschung (BMBF),
Deutsche Forschungsgemeinschaft (DFG),
Helmholtz Alliance for Astroparticle Physics (HAP),
Initiative and Networking Fund of the Helmholtz Association,
Deutsches Elektronen Synchrotron (DESY),
and High Performance Computing cluster of the RWTH Aachen;
Sweden {\textendash} Swedish Research Council,
Swedish Polar Research Secretariat,
Swedish National Infrastructure for Computing (SNIC),
and Knut and Alice Wallenberg Foundation;
European Union {\textendash} EGI Advanced Computing for research;
Australia {\textendash} Australian Research Council;
Canada {\textendash} Natural Sciences and Engineering Research Council of Canada,
Calcul Qu{\'e}bec, Compute Ontario, Canada Foundation for Innovation, WestGrid, and Digital Research Alliance of Canada;
Denmark {\textendash} Villum Fonden, Carlsberg Foundation, and European Commission;
New Zealand {\textendash} Marsden Fund;
Japan {\textendash} Japan Society for Promotion of Science (JSPS)
and Institute for Global Prominent Research (IGPR) of Chiba University;
Korea {\textendash} National Research Foundation of Korea (NRF);
Switzerland {\textendash} Swiss National Science Foundation (SNSF).
We acknowledge financial and infrastructure support by RWTH Aachen University and TU Dortmund University.

    \bibliographystyle{unsrtnat}
    \bibliography{Bibliography}

\end{document}